\documentstyle[pre,aps,epsfig,twocolumn]{revtex}
\begin{document}
\draft
\flushbottom
\twocolumn[
\hsize\textwidth\columnwidth\hsize\csname @twocolumnfalse\endcsname

\title{Classical projected phase space density of billiards 
and its relation to the quantum Neumann spectrum} 
\author{Debabrata Biswas}
\address{
Theoretical Physics Division \\
Bhabha Atomic Research Centre \\
Mumbai 400 085, INDIA}
\date{\today}
\maketitle

\begin{abstract}
A comparison of classical and quantum evolution usually
involves a quasi-probability distribution as a quantum analogue
of the classical phase space distribution. In an alternate approach that 
we adopt here, the classical density is projected 
on to the configuration space. We show that for billiards, 
the eigenfunctions of the coarse-grained projected 
classical evolution operator are identical to a first approximation 
to the quantum Neumann eigenfunctions. However, even though there exists a 
correspondence between the respective eigenvalues,
their time evolutions differ. This is demonstrated numerically 
for the stadium and lemon shaped billiards.
\end{abstract}

\vskip 0.15 in
%\pacs{PACS number(s): 05.45.Mt, 03.65.Sq, 31.15.Gy,05.45.Ac}
\date{today}
]
\narrowtext
\tightenlines

%\pacs{PACS number(s): 05.45.Mt, 03.65.Sq, 05.45.Ac}
%date{today}
%\end{frontmatter}
\newcommand{\be}{\begin{equation}}
\newcommand{\ee}{\end{equation}}
\newcommand{\bea}{\begin{eqnarray}}
\newcommand{\eea}{\end{eqnarray}}
\newcommand{\Lop}{{\cal L}}
\newcommand{\DB}[1]{\marginpar{\footnotesize DB: #1}}
\newcommand{\q}{\vec{q}}
\newcommand{\kt}{\tilde{k}}
\newcommand{\Lopn}{\tilde{\Lop}}

A comparison of classical and quantum dynamics in terms of
appropriate evolution operators is of much 
interest \cite{sano,braun,haake,nonnenmacher}.
Besides improving our understanding of the statistical
properties of quantum spectral fluctuations, these studies are 
expected to shed light on the effective 
irreversibility (e.g. relaxation to the invariant density) 
observed in low-dimensional isolated systems which 
are otherwise time reversible.
 
In the probabilistic approach to dynamics, classical time evolution 
is studied in terms of the propagation of the phase space density, 
$\rho(p,q)$. The evolution of $\rho(q,p)$  is governed by the 
Perron-Frobenius (PF) operator, $\Lop^t$, 

\be
\Lop^t \circ \rho({\bf x}) = \int \delta({\bf x} - {\bf f}^t({\bf x}')) 
\rho({\bf x}') d{\bf x'} \label{eq:defPF}
\ee

\noindent
where ${\bf x} = ({\bf q},{\bf p})$ is  a point in phase space 
and ${\bf f}^t({\bf x} )$ is its position at time $t$.
In the Hilbert space of phase space functions, $\Lop^t$ is
unitary \cite{lasota,gaspard}. Thus, the eigenvalues lie on
the unit circle. The existence of an invariant density, $\rho_0$, 
implies $\Lop^t \circ \rho_0 = \rho_0$. A system is
ergodic if the unit eigenvalue is nondegenerate. 
A knowledge of the spectral decomposition 
of $\Lop^t$ allows one to evaluate correlations, 
averages and other quantities
of interest.  For 
integrable systems, the spectrum of $\Lop^t$ is discrete
while for mixing systems, there is a continuous spectrum 
apart from the unit eigenvalue. The decay to the
invariant density is connected to the continuous part of
the spectrum. When the decay is exponential as in case of a class of
chaotic systems, the fourier transforms of time-correlations 
have poles in the complex frequency plane or broad peaks 
(resonances) along the real frequency axis, the positions
of which are independent of the observable chosen \cite{ruelle}.

A starting point for the comparison of classical and
quantum dynamics is usually a quasi-probability distribution 
involving the density operator, $\hat{\rho} = |\psi><\psi|$.
This enables one to ``lift'' the quantum state
to the phase space. The process is however not unique and 
depends on the ordering scheme used to arrange the 
non-commuting operators $(\hat{q},\hat{p})$ \cite{puri,zhang}.
A commonly used quasi-probability distribution is the Husimi function
which is a coherent (most classical) state 
representation of the quantum density operator. 
The Husimi propagator is 
thus a quantum analogue of the Perron-Frobenius operator
and the eigenstates of the two provide a means of comparing
quantum and classical evolution.

While the exact Husimi function does not decay to the 
invariant density, it is found that coarse graining of the
phase space leads to a loss of unitarity  \cite{haake,nonnenmacher}. 
For the kicked top \cite{haake} and maps on the 
torus \cite{nonnenmacher}, the eigenvalues of the coarse grained
classical and quantum propagator 
have been found to be identical.

The primary aim of this paper is to explore the existence of 
a correspondence between the classical and quantum spectrum
for the class of  systems referred to as billiards.
However, rather than ``lifting'' the quantum description
to the phase space using a quasi-probability distribution, 
we shall project the classical
density to the configuration space by integrating out the momentum:
$\rho({\bf q})=\int \rho({\bf q},{\bf p})~d{\bf p}$.
While we shall restrict ourselves to billiards here, the
essential idea of seeking a quantum-classical correspondence
using the projected density instead of the full phase space
density is applicable to other systems as well. In the following,
we shall  show that for billiards, 
there exists a correspondence between the
eigenvalues  of the coarse-grained projected 
classical evolution operator ($\Lop_P^t$) and the quantum 
Neumann spectrum,
while the respective eigenfunctions are identical to a first 
approximation.
However, the eigenvalues evolve differently with time so that
classical and quantum evolution differ \cite{dirichlet} 
 
Apart from being a paradigm in the field of
classical and quantum chaos, billiards have relevance in variety
of contexts. The Helmholtz equation describing the quantum
billiard problem also describes acoustic
waves, modes in microwave cavities and has relevance
in studies on ``quantum wells'', ``quantum corrals'', mesoscopic 
systems and nanostructured materials.

In a classical billiard, a particle moves freely inside a given
enclosure and reflects specularly from the wall at the boundary.
Depending on the shape of the boundary, billiards exhibit the
entire range of behaviour observed in other dynamical systems.
They also provide a means of coarse graining 
that is perhaps unique. The dynamics of billiards with smooth 
boundaries can be coarse grained by polygonalizing the 
boundary \cite{ford,poly,arb3}. Rational polygonal billiards
are non-ergodic and non-mixing. However, the short time 
dynamics of a polygonalized billiard can approximate that of
the smooth billiard  \cite{ford}.

The quantum billiard problem consists of determining the eigenvalues and
eigenfunctions of the Helmholtz equation 
$\nabla^2 \psi(q) + k^2 \psi(q) = 0$
with $\psi(q) = 0$ (Dirichlet boundary condition) or 
$\hat{\bf n}.\nabla\psi = 0$ (Neumann boundary condition; $\hat{\bf n}$
is the unit normal)
on the billiard boundary. Its semiclassical description holds the
key to the quantum-classical correspondence. However, since a 
quantum state (or the quasiprobability distributions constructed
out of it) can essentially resolve phase space structures of the size
of a Plank cell, polygonalization provides just as much information
about the quantum state at the semiclassical level \cite{poly,fnote0}.

We shall consider a polygonalized billiard as an 
example of a coarse-grained system.
The ``unfolded'' dynamics of a polygonalized billiard can be
viewed locally as a  straight line on a singly connected  
invariant surface consisting of 
multiple copies of the enclosure glued together appropriately
at the edges, each copy denoting a momentum direction
that is related to the previous one by the law of reflection
at the glued edge (for a figure, see \cite{arb2,cap}). 
As the magnitude of the momentum ($p$) and
the angle $\varphi$ that ${\bf p}$ makes with (say) the X-axis
are conserved, it is convenient to treat ${\bf p}$ in polar 
coordinates ($p,\varphi$). 
Transforming from ($p_x,p_y$) to ($p,\varphi$), the ${\bf p}$ 
integration in eq.~(\ref{eq:defPF}) simplifies as

\bea
& \; & \int \tighten dp_x dp_y \delta(p_x - {p_x'}^t({\bf q'},{\bf p'}) 
\delta(p_y - {p_y'}^t({\bf q'},{\bf p'}) h({\bf q'},{\bf p'}) \nonumber \\
& ~ & ~~~~=  
\int d\varphi~\delta(\varphi - \varphi') h({\bf q'_u},\varphi';p) 
= h({\bf q'_u},\varphi;p) 
\eea

\noindent 
where ${\bf q_u}$ is a point on the unfolded space, 
${p_x'}^t({\bf q'},{\bf p'})$ (${p_y'}^t$) is the
x (y) component of the momentum at time $t$ for the initial phase
space coordinate $({\bf q'},{\bf p'})$ and 

\be
h({\bf q'_u},\varphi;p) = \delta({\bf q_u} - {\bf q_u}'^t({\bf q_u}';
\varphi,p)) \rho({\bf q_u'};\varphi,p)
\ee

\noindent
Thus

\be
\Lop^t \circ \rho  =  \int d{\bf q_u'}  
\delta({\bf q_u} - {\bf q_u}'^t({\bf q_u'};\varphi,p)) \rho({\bf q_u'}).
\ee

\noindent
Note that the time evolution of $\rho$ depends on $\varphi$ 
through the kernel as ${\bf q_u}'^t$ depends on both the initial
position and momentum. Thus $\Lop^t$ = $\Lop^t(\varphi)$. 

Projection onto to the configuration space requires an 
integration over the angle $\varphi$. The time evolution of the
projected density is thus given by:

\bea
\Lop_{P}^t \circ \rho ({\bf q}) & = & 
{1 \over 2\pi} \int_0^{2\pi} d\varphi~ \Lop^t(\varphi) \circ \rho({\bf q}) 
\nonumber \\
& = & {1 \over 2\pi} \int d{\bf q_u}' d\varphi~ 
\delta({\bf q_u} - {\bf q_u}'^t({\bf q_u}';p,\varphi)) \rho({\bf q}')  
\label{eq:projected}
\eea

The spectrum of  $\Lop_P^t$, can be studied by evaluating its trace

\bea
{\rm Tr}~ \Lop_P^t & = & {1 \over 2\pi} \int d\varphi \sum
e^{\lambda_n(\varphi)t} \\
& = & {1 \over 2\pi} \int d{\bf q_u} 
\int d\varphi~ \delta({\bf q_u} - {\bf q_u}'^t({\bf q_u}';p,\varphi)).
\eea

\noindent
Note that
due to the multiplicative nature of the Perron-Frobenius 
operator, $\Lop^t$, its eigenvalues 
$\Lambda_n(t;\varphi)$ are of the form $e^{\lambda_n(\varphi)t}$. 
It can be shown that for $t > 0$ \cite{arb1,arb2}

\be
{\rm Tr}~ \Lop_P^t  \simeq  
 N C +  N \sum_{n=1}^{\infty} g(\sqrt{E_n}l) \label{eq:pre_final}
\ee 

\noindent
where $\{E_n\}$ refers to the quantum Neumann spectrum, $l = tv$
where $v$ refers to the velocity,
$g(x) = \sqrt{2 /(\pi x)}~{\cos}(x - \pi/4)$, $N$ is the maximum
number of allowed momentum directions 
and $C \simeq 1/N$ is a constant \cite{prl1,rapid1}. Since
$g(x) \simeq {1\over 2\pi}\int_0^{2\pi}~e^{ix\sin(\varphi)}d\varphi$
for large $x$, it follows that for $v = 1$,
 
\be
\lambda_n(\varphi)= i\sqrt{E_n}\sin(\varphi) \label{eq:final}
\ee             

\noindent
Thus, the power spectrum of a projected density contains peaks
at $ \sqrt{E_n}$. 

The above correspondence between the classical and quantum 
spectrum of polygonal billiards arises from a similarity in 
the traces of $\Lop_P^t$ and the quantum propagator when expressed 
in terms of periodic orbits. The correspondence however
strictly holds for large $E_n$ 
so long as one uses a delta function kernel 
in $\Lop^t$ \cite{arb2}. For smaller value of $E_n$, 
the correspondence exists if the delta function kernel in 
eq.~(\ref{eq:projected}) is smoothened \cite{cap,arb2}. 
This effectively results in a coarse graining
of the dynamics and in generic cases, leads to the 
the inclusion of a higher order term
in the classical and quantum trace \cite{arb2,cap}.
Despite the coarse graining, $\{E_n\}$ in most cases refers to the 
{\em approximate} quantum Neumann spectrum as the semiclassical 
trace formula, which is used in arriving at the correspondence, 
remains inexact. 
For integrable polygons such as the rectangle, the
correspondence between the spectrum of the projected
Perron-Frobenius operator and the Neumann spectrum is exact and 
can be shown directly \cite{arb2}.

Note that despite the correspondence,
quantum time evolution differs from evolution due to $\Lop_P^t$
as the eigenvalues evolve differently.
The quantum Neumann eigenfunctions are however
approximate eigenfunctions of $\Lop_P^t$. 
This has been established \cite{arb3} for a quasiclassical 
adaptation of $\Lop_P^t$ when the Dirichlet 
eigenstates are of interest.
For the Neumann problem, a similar derivation follows provided  
the quasiclassical kernel is replaced by the classical kernel. 
We shall demonstrate numerically that $\psi_n({\bf q})$ 
indeed approximates the quantum Neumann eigenfunctions.

For generic polygons, the unit  
eigenvalue of $\Lop^t$ is nondegenerate 
and the corresponding eigenfunction is a constant.
This is true as well for the projected operator, $\Lop_P^t$.
Thus $\lambda_0 = 0$. The quantum Neumann problem
also has a constant as its ground state eigenfunction
and the corresponding eigenenergy $E_0=0$. This is consistent
with the results presented here. 

In order to determine the eigenvalues and  eigenfunctions of $\Lop_P^t$, 
we shall 
first evaluate its smoothened kernel

\bea \nonumber
K_{P}({\bf q},{\bf q}',t) & = & {1\over 2\pi} 
\int~d\varphi~\delta_\epsilon({\bf q} - {{\bf q}'}^t(\varphi))
 \\ & = & \sum_n \psi_n({\bf q}) {\psi_n}^*({\bf q}')
\Lambda_n(t) \label{eq:smoothen}
\eea

\noindent
as a function of time. Here $\delta_\epsilon$ is a smoothened delta 
function and $\psi_n({\bf q})$ are the eigenfunctions of the
projected Perron-Frobenius operator, $\Lop_P^t$. As an example of
the smoothened delta function, we consider the hat function
which is zero outside a cell of size $\epsilon$ \cite{fnote_epsilon}. 
The $\varphi$ integration is performed by 
shooting trajectories from a point ${\bf q}'$ at various angles
and evaluating the fraction of trajectories in a 
cell of size $\epsilon$
at ${\bf q}$ \cite{arb2}.
Since $\lambda_n = i \sqrt{E_n} \sin(\varphi)$, for $v=1$, a fourier
transform of $K_P({\bf q},{\bf q}',t)$  has peaks at $k = \sqrt{E_n}$,
the width depending on $\epsilon$
and the heights on $\psi_n({\bf q})$. An eigenfunction corresponding
to a particular eigenvalue can thus be measured by varying ${\bf q}$
and measuring the height of the desired peak.

\begin{figure}[tbp]
{\vspace*{-.2in}}
\hspace*{-.75cm}\epsfig{figure=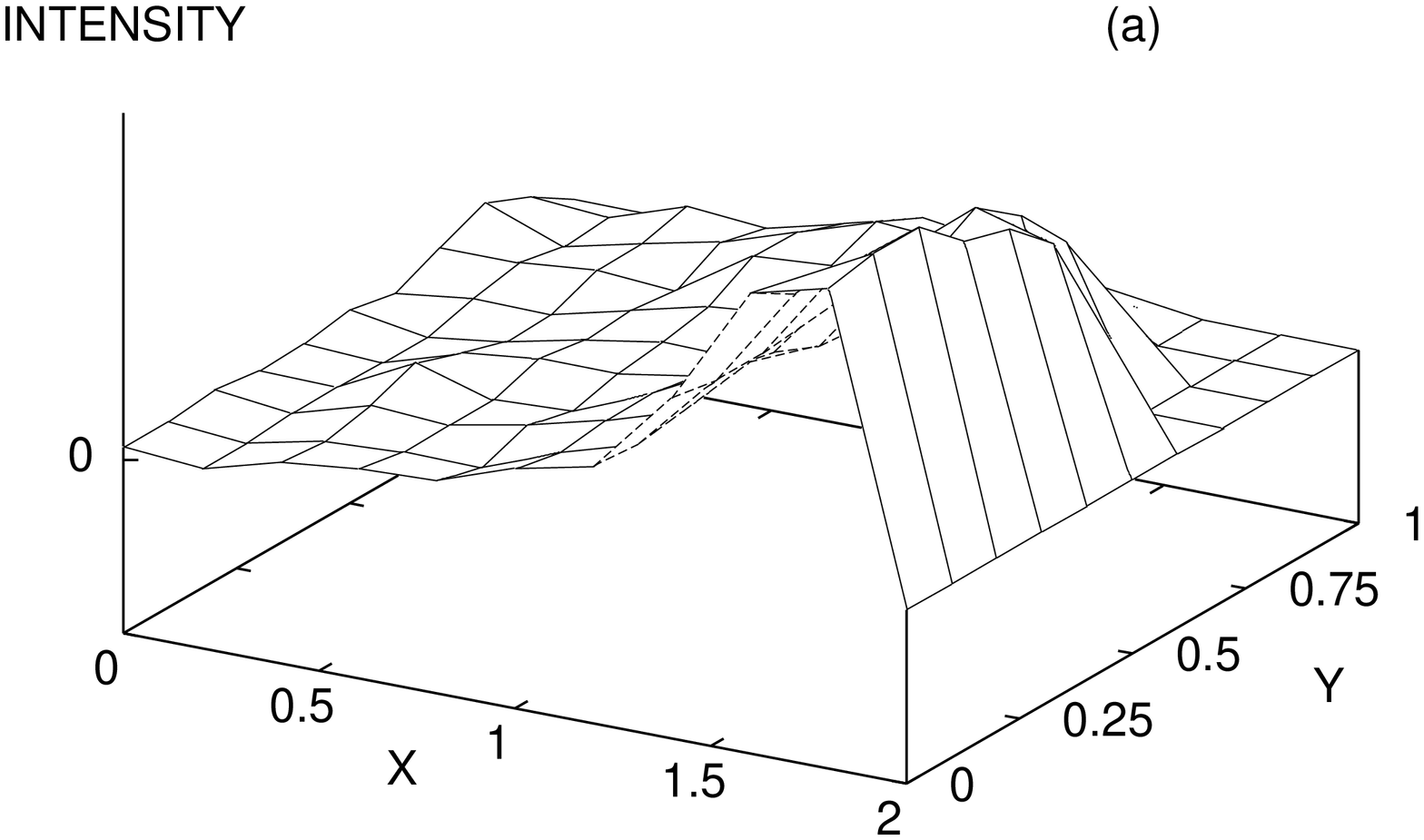,width=6cm,angle=270}
{\vspace*{-0.3in}}
\hspace*{-.5cm}\epsfig{figure=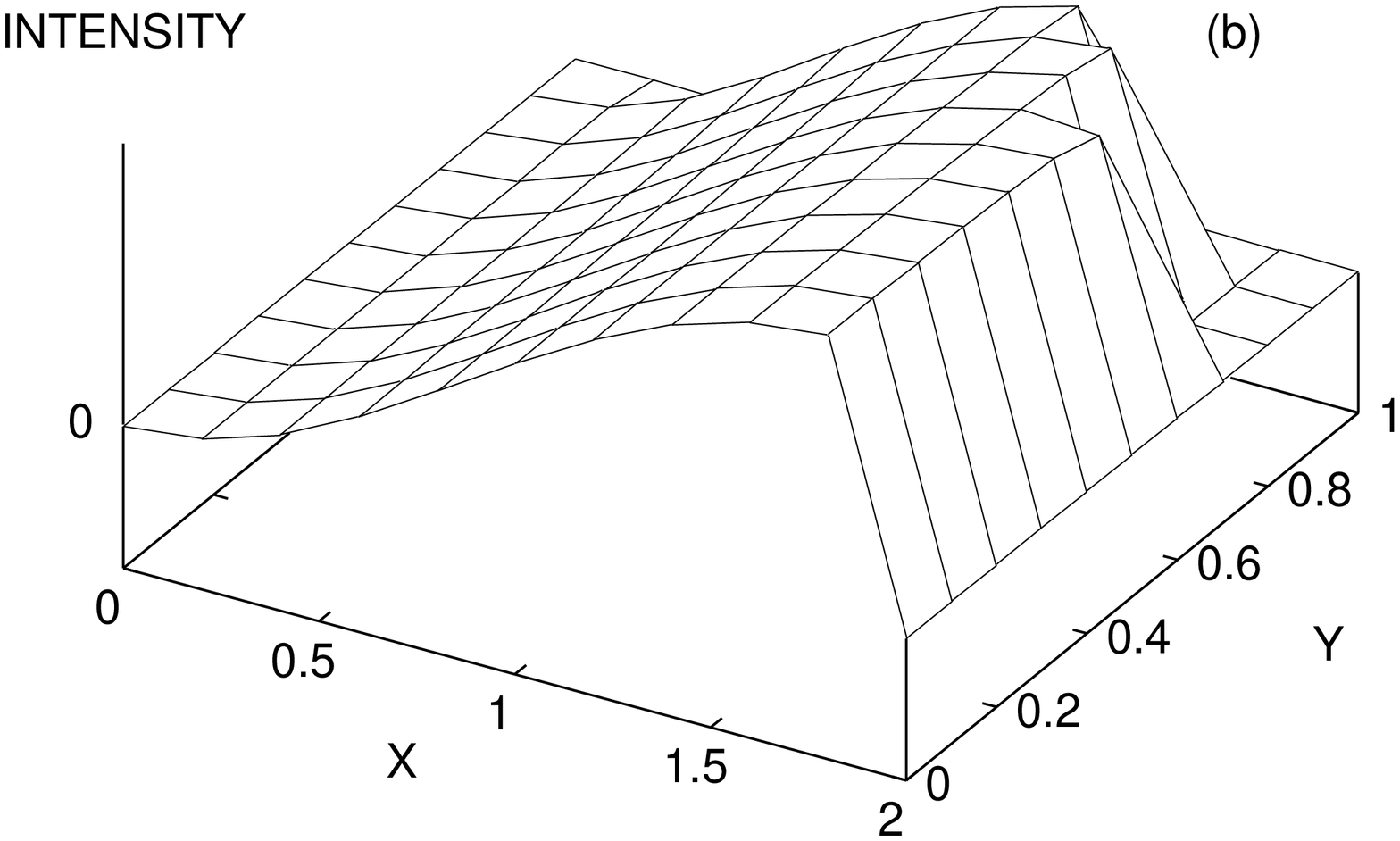,width=6cm,angle=270}
%{\vspace*{.2in}}
\caption[ty]{An eigenfunction of $\Lop_P^t$ corresponding to the 
first non-zero eigenvalue of the polygonalized  stadium.
It is symmetric in $Y$ (Neumann) and antisymmetric in $X$ (Dirichlet).
In the region outside the stadium, the value has been set to zero.  
(b) its quantum counterpart in the smooth stadium.}
\label{fig:1}
\end{figure}

We now present our numerical results for the stadium and lemon shaped
billiards. The chaotic stadium shaped  billiard 
that we consider, consists
of two parallel straight segments of length $2$ joined on either end
by a semicircle of unit radius. This has been polygonalized
using 10 segments to approximate the semicircle.
Figure~1a shows the eigenfunction of $\Lop_P^t$ 
corresponding to the first peak at a  non-zero $k$ ($k \simeq 0.70$) in
the power spectrum of (the smoothened) kernel $K_P({\bf q},{\bf q},t)$ 
for the polygonalized stadium. Only the first quadrant is shown
due to the reflection symmetry of the system. 
Note that we have plotted the
intensities as the peak heights are proportional to $|\psi_n(q)|^2$. 
Figure~1b shows the 
corresponding quantum Neumann eigenfunction of the smooth
stadium at $k \simeq 0.87$ found using the boundary integral
technique.
The eigenfunction of the projected 
Perron-Frobenius operator clearly approximates the quantum
Neumann eigenfunction. This is true for 
other stadium eigenfunctions as well.

\begin{figure}[tbp]
{\vspace*{-0.2in}}
\hspace*{-0.75cm}\epsfig{figure=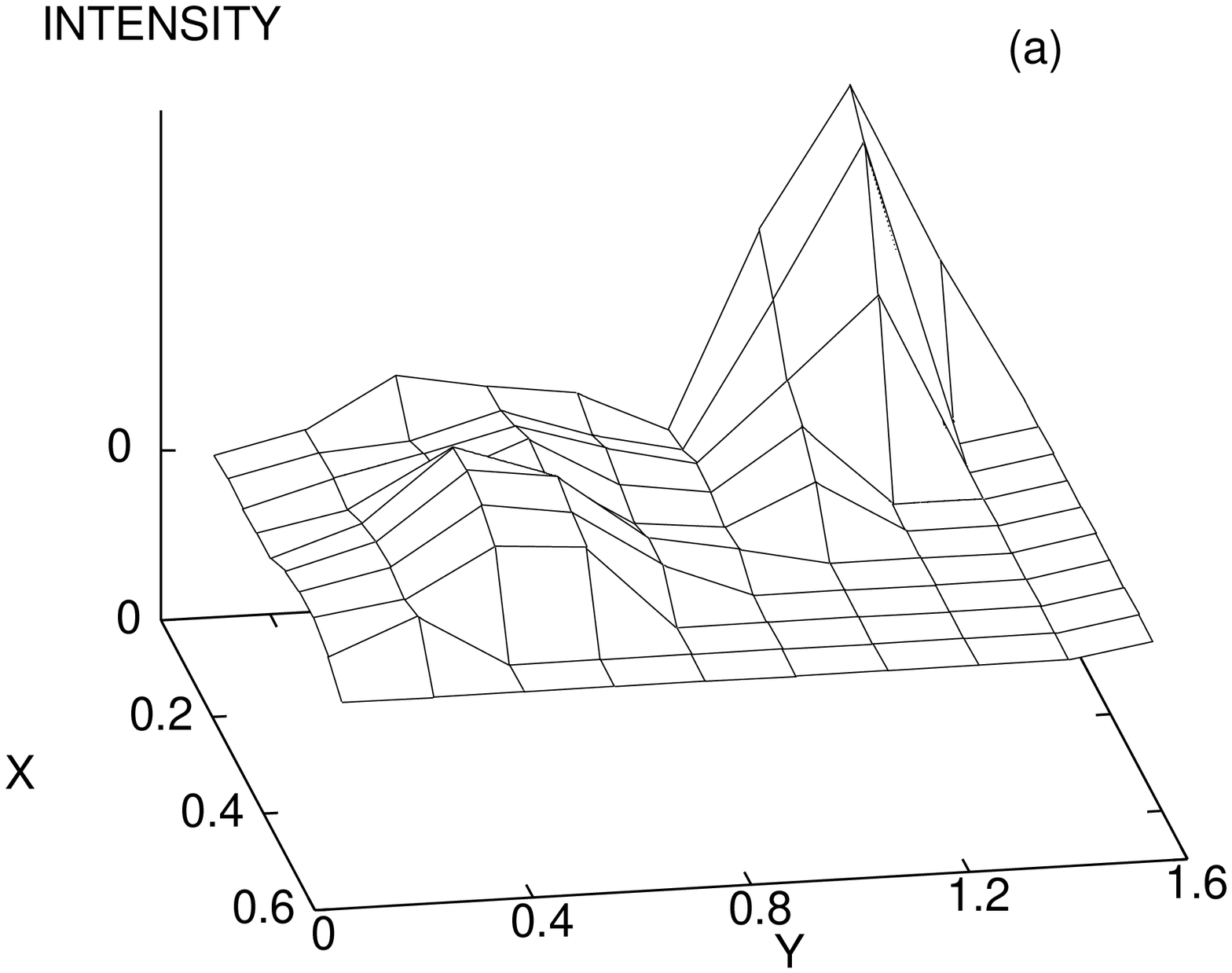,width=6cm,angle=270}
{\vspace*{-0.3in}}
\hspace*{-0.5cm}\epsfig{figure=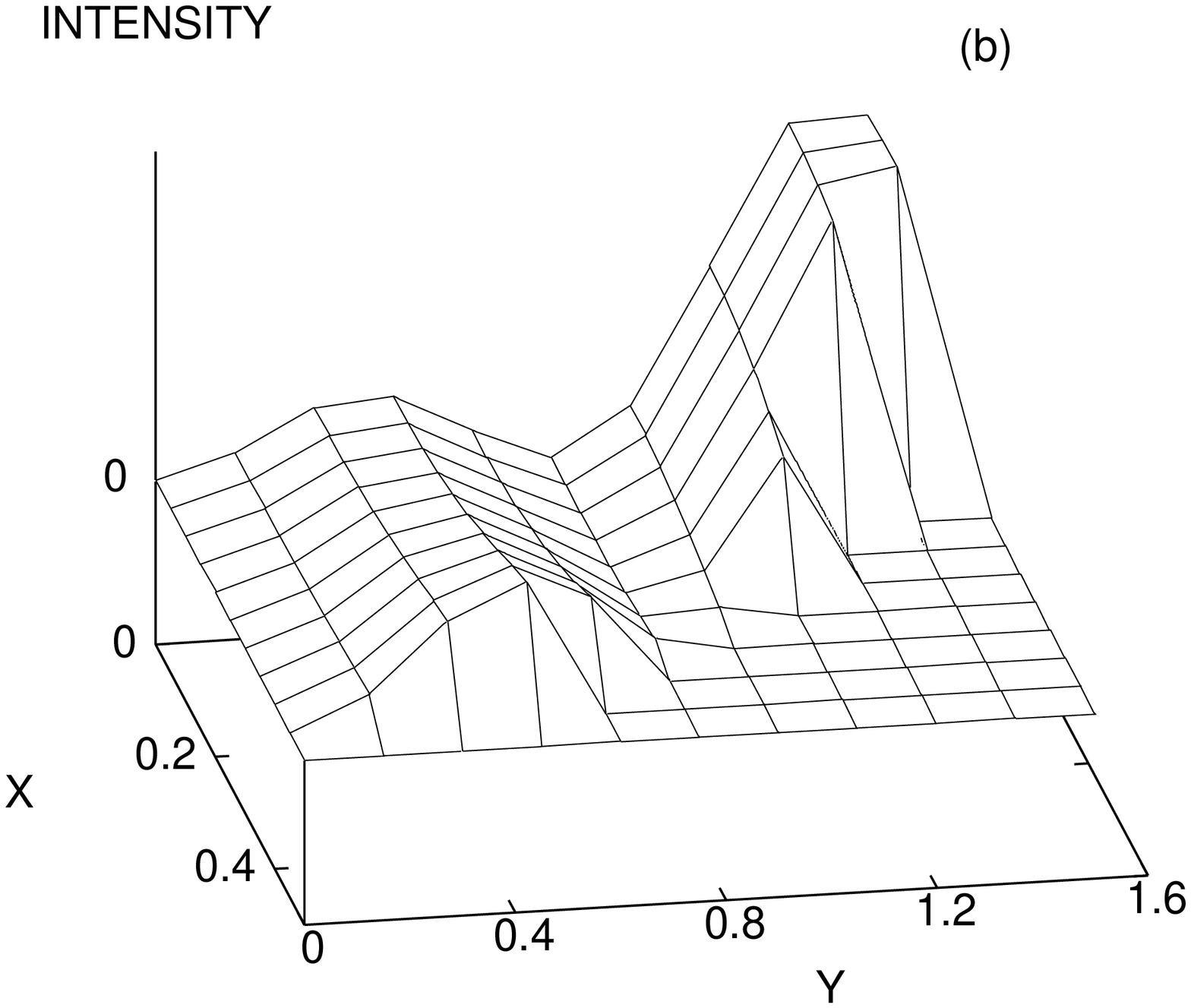,width=6cm,angle=270}
%{\vspace*{.7in}}
\caption[ty]{(a) An eigenfunction of $\Lop_P^t$ corresponding to the 
third non-zero eigenvalue of the polygonalized lemon shaped billiard.
It is symmetric in $X$ and antisymmetric in $Y$. 
(b) its quantum counterpart in the smooth lemon with $k \simeq 3.51$.}
\label{fig:2}
\end{figure}

We consider next a lemon shaped enclosure constructed by
the intersection of two circles of radius 2.5 
centred at (2,0) and (-2,0) respectively.
Each arc is approximated by seven segments. 
Figure 2 shows a comparison similar to fig.~1 for this enclosure
for the third peak in the power spectrum of $K_P$ at $k \simeq 3.08$. 
The agreement with the exact quantum eigenfunction ($k \simeq 3.51$) 
of the smooth lemon shaped billiard is again reasonable.

We have thus seen that there is a correspondence, albeit
approximate, between the eigenstates of the quantum
and projected classical evolution operators. 
The theoretical basis (see also \cite{arb1,arb2}) 
clearly indicates that the eigenstates obtained using
this method are at best ``semiclassical'' in nature since
the quantum trace formula, which is necessary in order 
to connect the eigenvalues of $\Lop_{P}$ with the quantum 
Neumann eigenvalues, is only approximate
as higher order corrections  have been neglected. In cases
where the corrections are zero (such as in the rectangular
or equilateral billiards), the quantum Neumann eigenstates
are identical to the eigenstates of $\Lop_{P}$ and the
smoothening parameter $\epsilon$ can be made small.
In general, at higher energies, 
the peaks in the fourier transform of 
$K_P({\bf q},{\bf q}',t)$ are generally 
harder to resolve as the density of eigenvalues increases with $k$.
Thus, measuring projected eigenfunctions
using trajectories becomes harder.

\begin{figure}[tbp]
%{\vspace*{-0.2in}}
\begin{center}
\hspace*{-0.75cm}\epsfig{figure=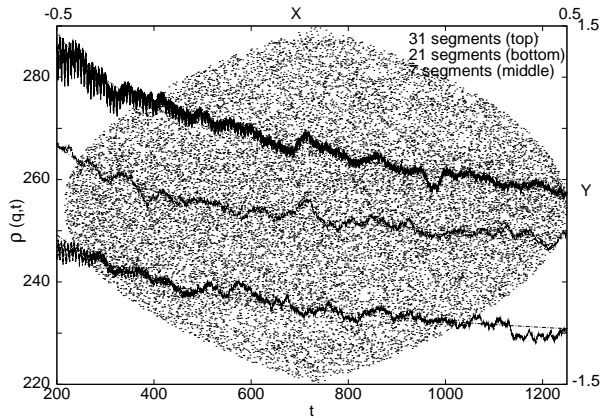,width=6cm,angle=270}
\end{center}
\caption[ty]{The time evolution of the locally space and time averaged 
configuration space density, $\rho({\bf q,t})$, in a lemon shaped 
billiard with different degrees of polygonalization. 
The number of linear segments approximating
each arc is shown in the figure. The point 
${\bf q}$ is different in each case. The background shows
the density at $t \simeq 1250$ for the 31 segment case. 
}
\label{fig:3}
\end{figure}

Note that there
are basic differences in the interpretations of 
the quantum and projected classical 
eigenstates. Each quantum eigenfunction
is associated with a density which is invariant in time
while only the constant eigenfunction of $\Lop_P^t$ qualifies
as a configuration space density and is invariant. 
All other densities, $\rho({\bf q})$, 
decay to the invariant density on evolving with $\Lop_P^t$.
This is true even when the dynamics is integrable.
The eigenvalues of $\Lop_P^t$ thus form a decay spectrum.
As an example of the decay to the uniform density in polygonalized
billiards, we present in fig.~\ref{fig:3}, the evolution of
a localized projected phase space density, locally averaged in space and
time, for three different versions of the polygonalized lemon-shaped
billiard considered above. The time evolution of a projected
density can be expressed in terms of the eigenstates of $\Lop_P^t$ as

\be
\rho({\bf q},t) \simeq \rho_{av} +  {1 \over t^{1/2}} 
\sum_{n=1}^{\infty} c_n \psi_n({\bf q}) \cos(\sqrt{E_n}t - \pi/4)  
\ee

\noindent
for $t$ sufficiently large. The coefficients, \{$c_n$\}, 
apart from some factors, depends on 
$\int \psi_n({\bf q}) \rho({\bf q},0) d{\bf q}$. The approach
to the uniform density, $\rho_{av}$, is highly oscillatory
and the overall decay rate may differ from $t^{-1/2}$ due to 
the sum of oscillatory terms. We have estimated the overall
decay by locally 
averaging $\rho({\bf q},t)$ over time.
For each of the three cases in fig.~\ref{fig:3}, the best fit
of $a + b/t^{1/2}$ is also shown. These results testify that
there is a decay to the uniform density in polygonalized
enclosures.

Finally, a discussion on coarse-graining is important
to understand the significance of the result. 
There are two levels at which this has been carried
out. First, the boundary has been polygonalized. 
This enables us to connect
the quantum and classical eigenvalues analytically.
As the number of segments can be increased to approximate
the smooth billiard arbitrarily well, it might be expected 
that the eigenvalues and eigenfunctions converge to those
of the smooth billiard in both the quantum and classical
case. We have verified this numerically for a few 
eigenstates of stadium and
lemon billiards. 
Thus coarse graining via polygonalization might be 
unimportant for observing the correspondence.
The second and more significant coarse graining of the
dynamics is related to the smoothening of the delta function
kernel. This enables us to connect the
classical and quantum eigenvalues even for smaller
values of $\lambda_n$ and is hence indispensable.

Finally, it is worth noting that the determination of
exact eigenstates of the Perron-Frobenius operator is
generally nontrivial while quantum states are
easier to determine. The usefulness of the results
presented here thus lies in using quantum states to
evolve classical configuration space densities.

\newcommand{\PR}[1]{{Phys.\ Rep.}\/ {\bf #1}}
\newcommand{\PRL}[1]{{Phys.\ Rev.\ Lett.}\/ {\bf #1}}
\newcommand{\PRA}[1]{{Phys.\ Rev.\ A}\/ {\bf #1}}
\newcommand{\PRB}[1]{{Phys.\ Rev.\ B}\/ {\bf #1}}
\newcommand{\PRD}[1]{{Phys.\ Rev.\ D}\/ {\bf #1}}
\newcommand{\PRE}[1]{{Phys.\ Rev.\ E}\/ {\bf #1}}
\newcommand{\JPA}[1]{{J.\ Phys.\ A}\/ {\bf #1}}
\newcommand{\JPB}[1]{{J.\ Phys.\ B}\/ {\bf #1}}
\newcommand{\JCP}[1]{{J.\ Chem.\ Phys.}\/ {\bf #1}}
\newcommand{\JPC}[1]{{J.\ Phys.\ Chem.}\/ {\bf #1}}
\newcommand{\JMP}[1]{{J.\ Math.\ Phys.}\/ {\bf #1}}
\newcommand{\JSP}[1]{{J.\ Stat.\ Phys.}\/ {\bf #1}}
\newcommand{\AP}[1]{{Ann.\ Phys.}\/ {\bf #1}}
\newcommand{\PLB}[1]{{Phys.\ Lett.\ B}\/ {\bf #1}}
\newcommand{\PLA}[1]{{Phys.\ Lett.\ A}\/ {\bf #1}}
\newcommand{\PD}[1]{{Physica D}\/ {\bf #1}}
\newcommand{\NPB}[1]{{Nucl.\ Phys.\ B}\/ {\bf #1}}
\newcommand{\INCB}[1]{{Il Nuov.\ Cim.\ B}\/ {\bf #1}}
\newcommand{\JETP}[1]{{Sov.\ Phys.\ JETP}\/ {\bf #1}}
\newcommand{\JETPL}[1]{{JETP Lett.\ }\/ {\bf #1}}
\newcommand{\RMS}[1]{{Russ.\ Math.\ Surv.}\/ {\bf #1}}
\newcommand{\USSR}[1]{{Math.\ USSR.\ Sb.}\/ {\bf #1}}
\newcommand{\PST}[1]{{Phys.\ Scripta T}\/ {\bf #1}}
\newcommand{\CM}[1]{{Cont.\ Math.}\/ {\bf #1}}
\newcommand{\JMPA}[1]{{J.\ Math.\ Pure Appl.}\/ {\bf #1}}
\newcommand{\CMP}[1]{{Comm.\ Math.\ Phys.}\/ {\bf #1}}
\newcommand{\PRS}[1]{{Proc.\ R.\ Soc. Lond.\ A}\/ {\bf #1}}


\begin{references}

\bibitem{sano} M.~Sano, \PRE{59}, R3795 (1999).

\bibitem{braun} D.~Braun, Chaos {\bf 9} 730 (1999).

\bibitem{haake} C.~Manderfeld, J.~Weber and F.~Haake,
J.~Phys. A {\bf 34}, 9893 (2001).  

\bibitem{nonnenmacher} S.~Nonnenmacher, Nonlinearity {\bf 16}, 1685 (2003). 

\bibitem{lasota} A.~Lasota and M.~MacKey,
\newblock {\em Chaos, Fractals, and Noise; Stochastic Aspects of Dynamics},
\newblock Springer-Verlag, Berlin, 1994.

\bibitem{gaspard} P.~Gaspard, 
\newblock {\em Chaos, scattering and statistical Meachanics},
\newblock Cambridge University Press, 1998.

\bibitem{ruelle} M.~Pollicott, Invent. Math {\bf 81} 413 (1985); 
D.~Ruelle, \PRL{56}, 405 (1986)

\bibitem{puri} R.~R.~Puri,
\newblock {\em Mathematical Methods of Quantum Optics},
\newblock Springer, Berlin, 2001. 

\bibitem{zhang} W.~Zhang, D.~Feng and R.~Gilmore,
Rev. Mod. Phys. {\bf 62}, 867 (1990). 

\bibitem{dirichlet} For a correspondence between the 
eigenstates of a quasiclassical adaptation of $\Lop_P^t$ 
and the quantum Dirichlet spectrum, see \cite{arb2,arb3}.
Note however that $\Lop_P^t$ is not interpreted in \cite{arb2,arb3} 
as the evolution operator for a projected classical density. 

\bibitem{arb2} D.~Biswas, \PRE{63}, 016213 (2001).

\bibitem{arb3} D.~Biswas, \PRE{67}, 026208 (2003).

\bibitem{ford} J.~L.~Vega, T.~Uzer and J.~Ford, \PRE{48}, 3414 (1993).

\bibitem{poly} D.~Biswas, \PRE{61}, 5073 (2000).



\bibitem{fnote0} The degree of polygonalization depends on the
de Broglie wavelength, $\lambda$; see \cite{poly}.
 

\bibitem{arb1} D.~Biswas, {\tt chao-dyn/9804013} \& in {\em
Nonlinear Dynamics and Computational Physics}, ed. V.~B.~Sheorey,
Narosa, New Delhi, 1999.



\bibitem{prl1} D.~Biswas and S.~Sinha, \PRL{70},  916 (1993). 

\bibitem{rapid1} D.~Biswas, \PRE{54}, R1044 (1996). 


\bibitem{cap} D.~Biswas, \PRE{61}, 5129 (2000).

\bibitem{fnote_epsilon} The smoothening parameter 
$\epsilon \sim {\cal O}(1/k_{max})$ where 
$k_{max}$ is the largest eigenvalue of interest \cite{arb2}.   

%\bibitem{other_delta} Several other smoothened kernels can 
%be used; see \cite{arb2}.






\end{references}
\end{document}